\documentclass[aps,twocolumn]{revtex4-2}
\usepackage{amssymb}
\usepackage{natbib}
\usepackage{graphicx}
\usepackage{amsmath}
\usepackage[bookmarks = false,colorlinks = true,allcolors = blue]{hyperref}
\usepackage{bm}
\usepackage[all]{hypcap}
\usepackage{graphicx}
\usepackage{colortbl}
\usepackage{booktabs}
\usepackage{enumerate}
\usepackage{enumitem}
\usepackage{braket}
\usepackage{mathrsfs}
\usepackage{multirow}
\usepackage{upgreek}
\usepackage{textcomp}
\usepackage{MnSymbol}

\usepackage{lineno}
\usepackage{threeparttable}

\usepackage{array}
\newcommand{\PreserveBackslash}[1]{\let\temp=\\#1\let\\=\temp}
\newcolumntype{C}[1]{>{\PreserveBackslash\centering}p{#1}}
\newcolumntype{R}[1]{>{\PreserveBackslash\raggedleft}p{#1}}
\newcolumntype{L}[1]{>{\PreserveBackslash\raggedright}p{#1}}

\begin{document}
\title{Sequential generation of multiphoton entanglement with a Rydberg superatom}

\author{Chao-Wei Yang$^{1,\,2}$}
\author{Yong Yu$^{1,\,2}$}
\author{Jun Li$^{1,\,2}$}
\author{Bo Jing$^{3}$}
\author{Xiao-Hui Bao$^{1,\,2}$}
\author{Jian-Wei Pan$^{1,\,2}$}

\affiliation{$^1$Hefei National Laboratory for Physical Sciences at Microscale and Department
of Modern Physics, University of Science and Technology of China, Hefei,
Anhui 230026, China}
\affiliation{$^2$CAS Center for Excellence and Synergetic Innovation Center in Quantum
Information and Quantum Physics, University of Science and Technology of
China, Hefei, Anhui 230026, China}
\affiliation{$^3$Institute of Fundamental and Frontier Sciences, University of Electronic Science and Technology of China, Chengdu, China}

\maketitle

\textbf{
Multiqubit entanglement is an indispensable resource for quantum information science. In particular, the entanglement of photons is of conceptual interest due to its implications in measurement-based quantum computing~\cite{raussendorf2001one,raussendorf2003measurement,briegel2009measurement}, communication~\cite{munro2012,azuma2015all,zwerger2016measurement,borregaard2020one}, and metrology~\cite{gao2010d,liu2021j}. The traditional way of spontaneous parametric down-conversion already demonstrates entanglement of up to a dozen photons~\cite{zhong201812} but is hindered by its probabilistic nature. Here, we experimentally demonstrate an efficient approach for multi-photon generation with a Rydberg superatom, a mesoscopic atomic ensemble under Rydberg blockade~\cite{saffman2010quantum}. Using it as an efficient single-photon interface~\cite{saffman2002}, we iterate the photon creation process that gives rise to a train of temporal photonic modes entangled in photon number degree~\cite{nielsen2010b}. We detect the multiphoton entanglement via converting the photon number degree to a time-bin degree. Photon correlations verify entanglement up to 12 modes. The efficiency scaling factor for adding one photon is 0.27, surpassing previous results~\cite{schwartz2016deterministic,zhong201812,li2020multiphoton,istrati2020sequential}, and can be increased significantly without fundamental limitations. 
}

A most popular way of creating multiphoton entanglement is harnessing the process of spontaneous parameter down-conversion (SPDC)~\cite{pan2012multiphoton} or a similar one of spontaneous four-wave mixing. While each source emits one pair of entangled photons, and by engineering multiple sources, multiphoton entanglement can be created via interference and measurement~\cite{pan2012multiphoton}. Using this method, an entanglement of 12 photons has been achieved~\cite{zhong201812}. Nevertheless, the process of paired photon creation is probabilistic in nature, and the chance of success in each trial needs to be kept very low to limit the contribution of high-order events. Besides, the efficiency of the probabilistic fusion gate is intrinsically limited by a maximal value of 50\%. These issues result in a poor scaling as the photon number goes large. In addition, the number of entanglement sources used gets linearly increased as a function of photon number, which results in a very complex setup. 

A more efficient way is using a deterministic atom-photon interface (e.g. trapped neutral atom, ions, quantum dot) and creating multiple single photons in sequence~\cite{gheri1998,saavedra2000,schon2005,lindner2009}. Single photons get entangled via interaction with the atomic qubit in consecutive photon creations, and additional operations can manipulate the entanglement during the intervals. By using more atoms~\cite{economou2010} interacted in one setup, entanglement other than the one dimensional matrix product state~\cite{schon2005} can be created. These schemes are, in principle, deterministic and can scale up to a vast number of photons. Moreover, they have an additional advantage of being resource efficient since merely one setup is required. Albeit very promising in theory, it is technically demanding in experiments. An atom-cavity setup including other similar systems with strong coupling suits perfect for these schemes, but if the coupling strength is not high enough, the entanglement length will be rather limited. A previous experiment with a quantum dot reports the entanglement of three photons~\cite{schwartz2016deterministic}. Subsequent experiments improve to an entanglement of four photons~\cite{li2020multiphoton,istrati2020sequential} with the quantum dot as the photon sources, but the entangling operations are reverted to external photonic gates that are probabilistic. Meanwhile, similar experiments were reported in a circuit QED setup, and entangled microwave photons are created with an entanglement length limited by measurement noises~\cite{besse2020realizing}.   

In this paper, we address this scheme experimentally with a conceptually different setup of a mesoscopic atomic ensemble. We make use of the Rydberg blockade in the atomic ensemble to form a single superatom, which is a collective state with one atom in the Rydberg levels and other atoms in the ground state~\cite{saffman2010quantum}. Such a superatom can be prepared deterministically~\cite{dudin2012observation} and emits single photons in a well-defined mode via collective enhancement~\cite{saffman2002,dudin2012strongly}. We generate a train of temporal modes that are entangled in photon-number state. Afterwards, we convert the photon-number state entanglement to time-bin entanglement via combining two consecutive temporal modes as one time-bin qubit. Then we characterize the entanglements created via performing photon correlation measurements in different settings, which directly gives the fidelities. For modes of up to 12, the measured fidelity is clearly above the threshold of 0.5, proving a genuine GHZ entanglement~\cite{guhne2009entanglement}. We have also analyzed the limitations in our setup and made discussions on the efficiency scaling and fidelity scaling.

\begin{figure*}[htbp]	
	\centering
	\includegraphics[width=0.8\textwidth]{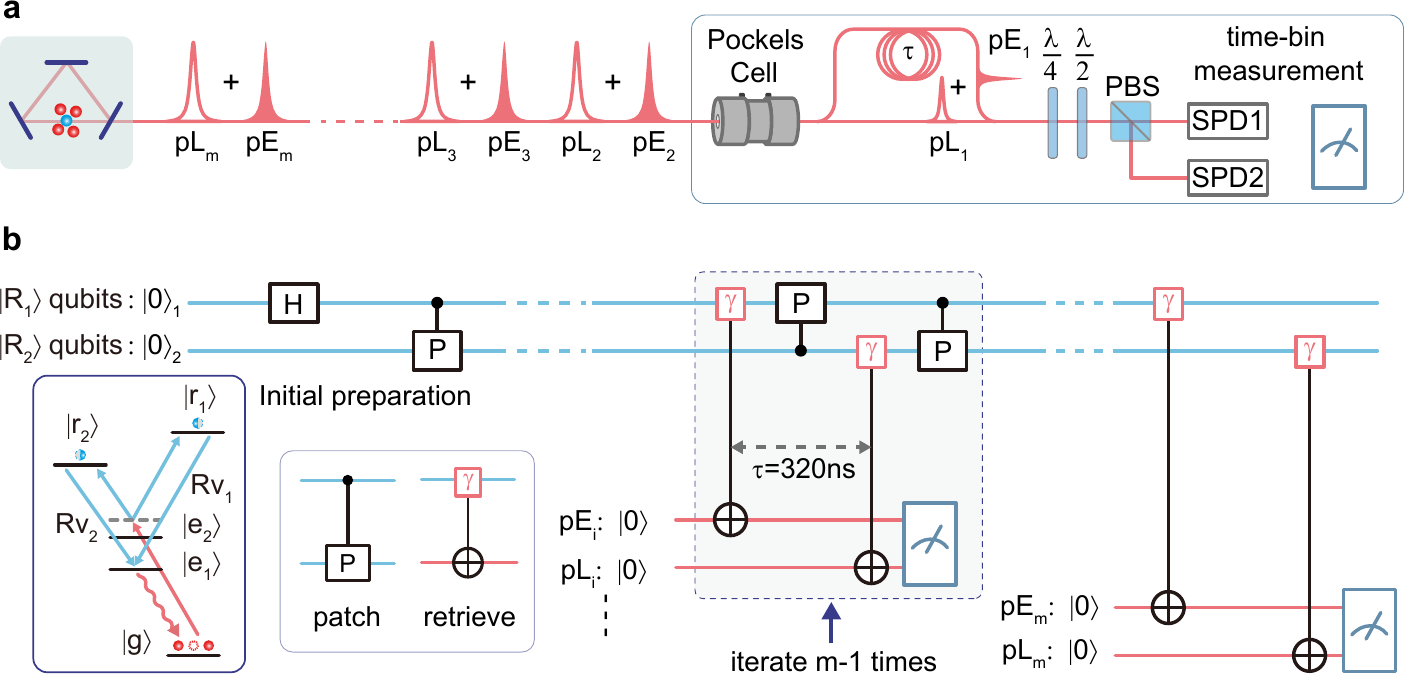}
	\caption{\textbf{Scheme for multi-photon entanglement generation.} \textbf{a}, Experimental setup. A mesoscopic atomic ensemble is harnessed for sequential photon generation. We make use of a ring cavity to enhance the retrieval process. Entanglement in the time-bin degree is verified with an unbalanced Mach–Zehnder interferometer, with a Pockels cell used to direct  different temporal modes to different interferometer arms. The photons are finally analyzed in polarization correlations. \textbf{b}, Circuit diagram and energy levels. Starting from the initial ground state $\ket{g}$, the superatom is prepared in an entangled state of $\ket{\varPsi}_a = (\ket{0}_1\ket{1}_2+\ket{1}_1\ket{0}_2)/\sqrt{2}$ via applying a collective Hadamard gate (H) for $\ket{r_1}$ and a patching $\pi$ pulse for $\ket{r_2}$. Afterwards, we perform iterated photon generations. In each iteration, we apply the retrieving and patching first for $\ket{r_1}$ and then for $\ket{r_2}$. The final atomic sates in $\ket{r_1}$ and $\ket{r_2}$ are also retrieved as an additional photon and which also disentangles the photon entanglement with the atomic state. $Rv_1$: retrieving pulse for $\ket{r_1}$, $Rv_2$: retrieving pulse for $\ket{r_2}$, $\mathrm{pE_j}$: the $j$-th photon in the \textit{early} mode, $\mathrm{pL_j}$: the $j$-th photon in the \textit{late} mode. }
	\label{fig:1setup}
\end{figure*}

The experimental scheme is shown in Fig.~\ref{fig:1setup} that simplifies from the original proposal~\cite{nielsen2010b} significantly. We make use of two Rydberg states $\ket{r_1}$ and $\ket{r_2}$, each of which hosts an atomic qubit defined as zero or one collective excitation in the form of $|R_{j}\rangle = N^{-1/2}\sum_{i=1}^{N}|g^{1}g^{2}\ldots r^i_{j}\ldots g^{N}\rangle$, with $\ket{g}$ being the atomic ground state, $N$ being the total number of atoms, $i$ being the atom index, and $j$ being the Rydberg state index. By applying a collective $\pi/2$ pulse coupling $\ket{g} \leftrightarrow \ket{r_1}$ followed with a collective $\pi$ pulse coupling $\ket{g} \leftrightarrow \ket{r_2}$, we first prepare an atomic entangled state of $\ket{\varPsi}_a = (\ket{0}_1\ket{1}_2+\ket{1}_1\ket{0}_2)/\sqrt{2}$. Afterwards, we apply the ``retrieving and patching'' sequence~\cite{sun2021} for $|r_1\rangle$, which converts the atomic number state to a photonic Fock state and recreate the atomic excitation again, resulting in a joint state of $(\ket{0}_1\ket{1}_2\ket{0}+\ket{1}_1\ket{0}_2\ket{1})/\sqrt{2}$, where a ket without a subscript denotes photonic state. Applying the same sequence for $\ket{r_2}$ will expand the entanglement to $(\ket{0}_1\ket{1}_2\ket{01}+\ket{1}_1\ket{0}_2\ket{10})/\sqrt{2}$. Iterating these processes $m-1$ times, we will arrive at an entanglement of $\ket{\varPsi}_{ap} = (\ket{0}_1\ket{1}_2\ket{01}^{\otimes m-1}+\ket{1}_1\ket{0}_2\ket{10}^{\otimes m-1})/\sqrt{2}$. Finally if we retrieve the two atomic qubits in sequence, a multiphoton entanglement of $2m$ temporal modes will be created in the form of $\ket{\varPsi}_m = (\ket{01}^{\otimes m}+\ket{10}^{\otimes m})/\sqrt{2}$, which is a GHZ-type entanglement.

Albeit a Rydberg superatom emits photons directionally, the photon collection efficiency gets limited by the optical depth. Improvement of the optical depth is challenging since the requirement of the Rydberg blockade imposes a limitation on the ensemble size. It is practical and efficient to use a low-finesse cavity to boost the collectively enhanced directional emission. As demonstrated in our previous work~\cite{yang2021single}, an overall efficiency (including preparing and fiber coupling) of up to 44\% has been achieved. Such an efficient single-photon interface forms the basis of conducting multiphoton experiments. In our setup, we capture hundreds of $^{87}$Rb atoms from the magneto-optical trap (MOT) with an 852 nm optical trap. The atoms are cooled to about 20 $\upmu$K with Doppler cooling and optical molasses. The excitation area is selected to be within the $1/e^2$ radius of about 6.5 $\upmu$m $\times$6.5 $\upmu$m $\times$1.4 $\upmu $m, which is much smaller than the blockade radius. The optical depth of the ensemble is about 1.9. In our experiment, the atoms are reloaded with a repetition rate of 6.7~Hz. Each loading takes 120 ms, followed with an entanglement generation phase of 30~ms in which we repeat multiphoton generation for 1000 cycles. In each cycle, the atoms are initialized in the ground state of $\ket{g}=\ket{5 S_{1/2}, F=2, m_F=2}$, and then excited to the Rydberg states via the intermediate state of $\ket{e_2}=\ket{5 P_{1/2}, F=2, m_F=1}$, and retrieved from $\ket{e_1}=\ket{5P_{1/2}, F=1, m_F=1}$ with a resonant read beam. To achieve fast and independent operations of two qubits, we make use of two Rydberg energy levels of $\ket{r_1}=\ket{91S_{1/2}, m_J=1/2}$ and $\ket{r_2}=\ket{91S_{1/2}, m_J=-1/2}$. The polarization and frequency are different for two coupling beams of 474 nm. And the Zeeman splitting of the Rydberg states is 2$\pi\times$15.4 MHz, with the bias magnetic field applied along the direction of photon retrieval.

\begin{figure*}[htbp]
	\centering
	\includegraphics[width=0.9\textwidth]{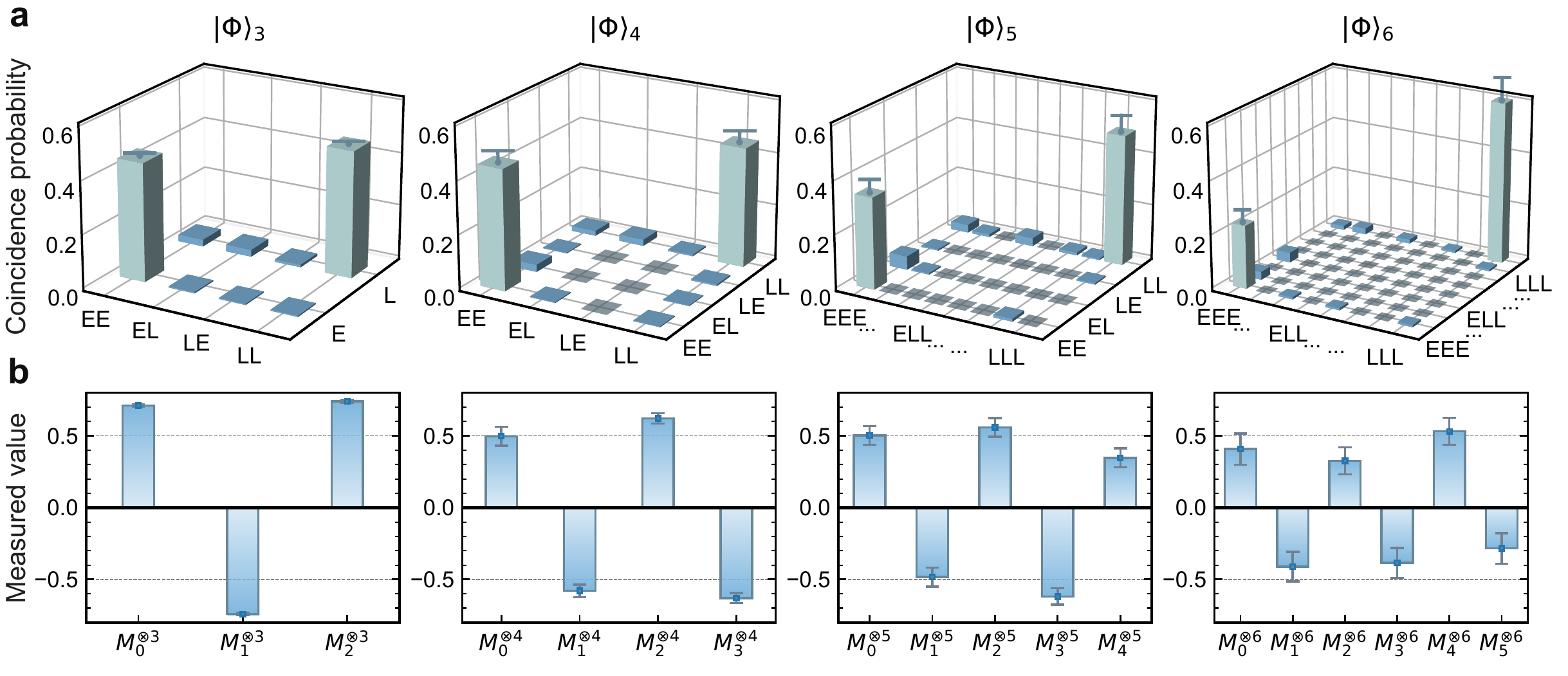}
	\caption{\textbf{Characterization of the multi-photon entanglements.} \textbf{a}, Measurement result in the eigen basis of $\ket{E}$/$\ket{L}$. The coincidence probabilities are calculated via normalization by the sum of all the $2^m$ m-fold coincidence counts. A gray bar indicates that the registered coincidence is zero. The left part of the labels denotes time-bin modes for the first $j$ photons, while the right part denotes time-bin modes for the remaining $m-j$ photons. For $m\geq5$, we omit showing several coincidence combinations due to limited space. \textbf{b}, Measurement result in the superposition bases of $M_n^{\otimes m}$. The error bars represent one standard deviation, deduced from the Poisson distribution of the detection of $m$-photon events. For the 6-photon entanglement, the overall coincidence count rate is about 17.2 per hour, and we set an average duration of 5.4 hours for measurement in each basis.}	
	\label{fig:maindata}
\end{figure*}

\begin{figure}[htbp]	
	\centering
	\includegraphics[width=0.9\columnwidth]{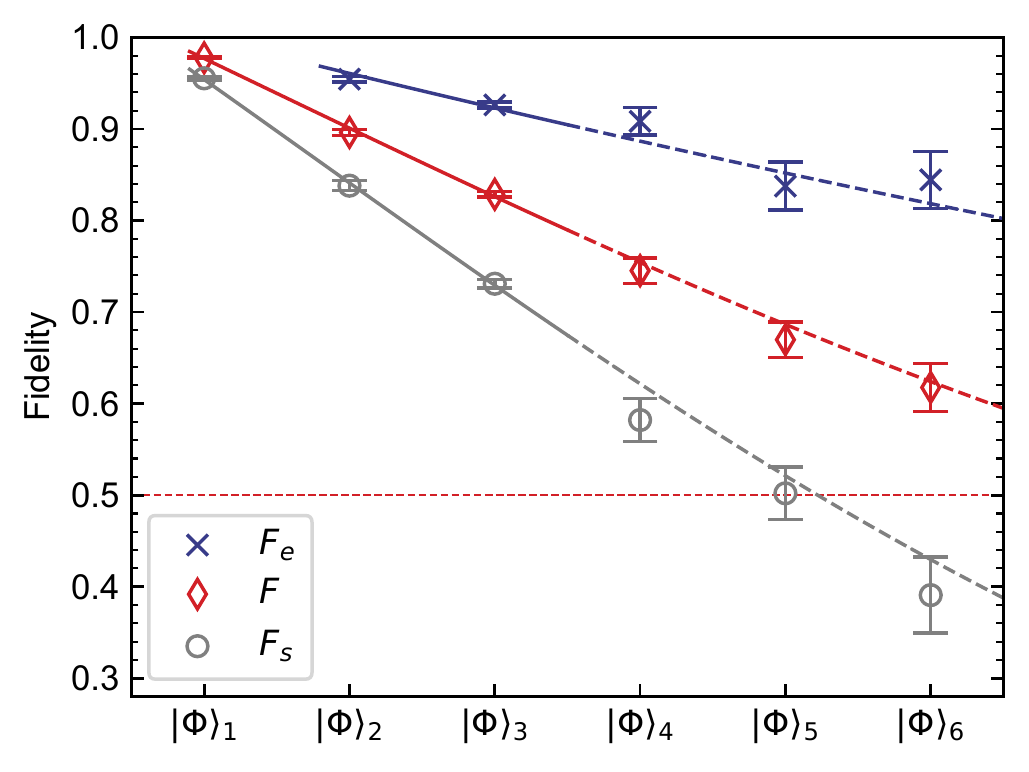}
	\caption{\textbf{Fidelities of the multi-photon entanglement.} The results from $\ket{\Phi}_3$ to $\ket{\Phi}_6$ are derived from the measurement results shown in Fig.~\ref{fig:maindata}. Original results on $\ket{\Phi}_1$ and $\ket{\Phi}_2$ are given in the Supplementary Information. The eigen basis fidelity $F_e$ is shown in blue, the superpositional basis fidelity $F_s$ is shown in gray, and the overall fidelity $F$ is shown in red. A dashed straight line of $F=0.5$ is plotted to mark the threshold to claim GHZ-type entanglement. Data points along the solid lines are used for curve fitting. }
	\label{fig:fideltiy}
\end{figure}

To verify the photon-number state entanglement, we combine two neighbouring temporal modes as a time-bin qubit via $\ket{10} \rightarrow \ket{E}$ and $\ket{01} \rightarrow \ket{L}$ where $E$ denotes an early mode and $L$ denotes a late mode. The multiphoton entanglement produced is thus converted to 
$\ket{\Phi}_m = (\ket{E}^{\otimes m}+\ket{L}^{\otimes m})/\sqrt{2}$. The time-bin qubits are measured with an unbalanced Mach-Zehnder interferometer. We utilize a Pockels cell to direct the early mode to the long arm, while 
direct the late mode to the short arm, thus enabling successive measurement in an arbitrary basis by rotating the waveplates, as shown in Fig.~\ref{fig:1setup}a. The relative time delay between the two arms is 320 ns, which matches the retrieval time difference between $\ket{r_1}$ and $\ket{r_2}$. In addition, we make use of an additional PBS and a Pockels cell (not shown) to minimize exposure of leakage from the excitation beam into the single photon detectors (SPD) by rapidly switching the noise pulses to the orthogonal polarization. To measure the fidelity of $m$-photon entanglement, we can decompose the fidelity operator into $m+1$ measurement settings:
\begin{equation}
\begin{split}
F=\frac{1}{2} (\ket{E}\bra{E}^{\otimes m} + \ket{L}\bra{L}^{\otimes m}) + \frac{1}{2m} \sum_{i=0}^{m-1} (-1)^i M_i^{\otimes m}
\end{split}
\end{equation}
with $M_i= {\rm cos}(i\pi/m)\sigma_x + {\rm sin}(i\pi/m)\sigma_y$ and $\sigma_{x,y}$ being the Pauli matrices. We perform a series of measurements from $m=1$ to $m=6$. In Fig.~\ref{fig:maindata}a, we give correlation measurement results in the eigen basis of $\ket{E}/\ket{L}$, which is later used to calculate the first part of fidelity as $F_e=\ket{E}\bra{E}^{\otimes m} + \ket{L}\bra{L}^{\otimes m}$. To estimate second part of the fidelity of $F_s= (1/m) \sum_{i=0}^{m-1} (-1)^i M_i^{\otimes m}$, we need further to perform measurement in a series of superpositional bases $M_i$. For each basis, we make the similar correlation measurements as Fig.~\ref{fig:maindata}a, and plot the calculated values of $M_i^{\otimes m}$ in Fig.~\ref{fig:maindata}b. For each entanglement, the final fidelity can be calculated as $F=(F_e+F_s)/2$. We summarize all the fidelities in Fig.~\ref{fig:fideltiy}, which drops gradually from a fidelity of 89.6\% $\pm$ 0.3\% for a two-photon Bell state $\ket{\varPhi}_2$ and 82.9\% $\pm$ 0.3\% for a three-photon GHZ sate to 61.8\% $\pm$ 2.6\% for a six-photon GHZ state. These results clearly verify that genuine GHZ-type entanglements~\cite{guhne2009entanglement} are produced, since the fidelities are well above the threshold of 50\%.

As shown in Fig.~\ref{fig:fideltiy}, the eigen basis fidelity $F_e$ drops exponentially as the photon number $m$ increases. By fitting the data with $F_e=\alpha^{m-1}$, we get an average factor of $\alpha=0.96$. Ideally, there should be only $\ket{E}^{\otimes m}$ and $\ket{L}^{\otimes m}$ components in Fig.~\ref{fig:maindata}a, but we observe clear contribution of coincidences in some special combinations. According to our analysis, there are two major imperfections that may happen during each iteration, i.e. the patching pulse fails to recreate an excitation, or an excitation experiences unexpected retrieval. As shown in the Supplementary Information, the patching pulse infidelity is estimated to be about 2\%, and may lead to a flip from $\ket{1}_1\ket{0}_2$ to $\ket{0}_1\ket{1}_2$, which we call a ``bit-flip'' error. This imperfection may also annihilate the Rydberg excitation and lead to $\ket{0}_1\ket{0}_2$, which will eliminate the generation of successive photons and shall not contribute to a $m$-fold coincidence. The unexpected retrieval probability is estimated to be 3\% for $\ket{R_1}$ and 1\% for $\ket{R_2}$.  For a positive flip from $\ket{1}_1\ket{0}_2$ to $\ket{0}_1\ket{1}_2$, both the patching pulse failure and unexpected retrieval will contribute, thus leading to a much higher flip probability. While for a negative flip from $\ket{0}_1\ket{1}_2$ to $\ket{1}_1\ket{0}_2$, only the unexpected retrieval contributes with the probability of about 1\%. These results agree well with the data in Fig.~\ref{fig:maindata}a, as we observe significantly higher contributions from the $\ket{E}^{\otimes k}\ket{L}^{\otimes m-k}$ terms than from the $\ket{L}^{\otimes k}\ket{E}^{\otimes m-k}$ terms. The mechanism also partly explains the relative reduction of $\ket{E}^{\otimes m}$ in comparison with $\ket{L}^{\otimes m}$ as $m$ goes larger.

The eigen basis fidelity $F_e$ sets an upper bound for the fidelity in the superpositional basis (see Supplementary Information). Further imperfections such as laser phase noise and interferometer phase stability will lead to a random phase between the two terms in $\ket{\Phi}_m$ that dephases the entanglement and lowers down the fidelity further~\cite{de2018analysis,levine2018high}. We perform independent measurements to characterize these phase noises. We get a standard deviation of $\sigma_{inter}=7.4^\circ$ for the interferometer phase fluctuation between the long arm and the short arm. Since the time scale of phase variance for the interferometer is significantly longer than the multiphoton generation sequence, we consider this noise as collective, which will lead to a phase scaling function of $m\sigma_{inter}$ for $m$ photons. For the phase noise from lasers, merely the fast components will contribute to the entanglement infidelity, since in each iteration in our entanglement generation sequence both the patching pulse and the read pulse are applied twice, making the slow components being cancelled out. The overall laser-induced phase fluctuation between the early and the late mode is measured to be $\sigma_{laser}=13^\circ$ defined as one standard deviation (see Supplementary Information for details). We consider this remaining laser noise as independent, which will result in a scaling of $\sqrt{m}\sigma_{laser}$ for $m$ photons. To evaluate how well our analysis describes the experiment, we fit the superpositional result in Fig.~\ref{fig:fideltiy} with a function of $F_s = F_e \,\beta_{\phi} \, \beta_{res}$,  with $\beta_{\phi}$ calculated from an overall standard deviation $\sqrt{m\sigma_{laser}^2+m^2 \sigma_{inter}^2}$ of the phase noise. Thus we get a residual factor of $\beta_{res} = 0.966^m$ which may come from the additional issues such as minor photon distinguishability between the two time-bin modes. 

The overall efficiency of detecting one photon is 9.4\%. After correcting the efficiencies of transmission ($\eta_t=50.8\%$) and detection ($\eta_d=68\%$), we will get an in-fiber efficiency of $\eta_f=27.2\%$ for each photon. This efficiency determines the probability for multiphoton generation via $\eta_m=\eta_f^m$, thus named as a scaling factor. In this definition, previous experiment with SPDC\cite{zhong201812} has a scaling factor of 0.16. And the experiments with quantum dot~\cite{istrati2020sequential,li2020multiphoton} still have a scaling factor of lower than 0.08. The single-photon efficiency $\eta_f$ is lower than the best value of 44\% we observed previously~\cite{yang2021single}. The reduction is mainly due to a prolonged interval between excitation preparation and retrieval, during which the collective excitation dephases slightly and the photonic retrieval efficiency gets reduced. 

In conclusion, we have demonstrated a new approach for multiphoton entanglement generation by making use of a Rydberg superatom. Two qubits are hosted in the superatom, and subjected to an alternating process of single-photon generation in sequence. Rydberg interaction between the two atomic qubits enables strong photon-number correlations in the retrieval temporal modes. We have verified GHZ-type entanglement of up to 12 modes. Our approach is advantageous since it is not only very resource-efficient by merely using single setup, but also shows much better scaling of adding more photons. In order to improve the efficiency scaling factor further, one may use an atomic ensemble with higher optical depth, a cavity with higher finesse and higher output coupling efficiency that may require placing the cavity inside the vacuum chamber. Further measures of improving the fidelity scaling factor includes using lasers with smaller phase noise,  eliminating retrieval cross talk, and improving the patching pulse fidelity etc. By incorporating more operations and more energy levels~\cite{nielsen2010b}, entanglements other than the GHZ type can be generated. Since the Rydberg states can be transferred to ground states for encoding and long-lived storage, our platform also has the potential of generating large hybrid entanglements with more atomic qubits and a large number of single photons, which may also enable new applications in quantum information science. 

This work was supported by National Key R\&D Program of China (No.~2017YFA0303902, No.~2020YFA0309804), Anhui Initiative in Quantum Information Technologies, National Natural Science Foundation of China, and the Chinese Academy of Sciences.

\bibliography{myref}

\clearpage

\setcounter{figure}{0}
\setcounter{table}{0}
\setcounter{equation}{0}

\onecolumngrid

\global\long\def\theequation{S\arabic{equation}}
\global\long\def\thefigure{S\arabic{figure}}
\renewcommand{\thetable}{S\arabic{table}}

\newcommand{\msection}[1]{\vspace{\baselineskip}{\centering \textbf{#1}\\}\vspace{0.5\baselineskip}}

\renewcommand{\thesubsection}{\arabic{subsection}}

\msection{SUPPLEMENTAL MATERIAL}

\section{More details on experimental setup}
\begin{figure}[hbtp]	
	\centering
	\includegraphics[width=0.6\textwidth]{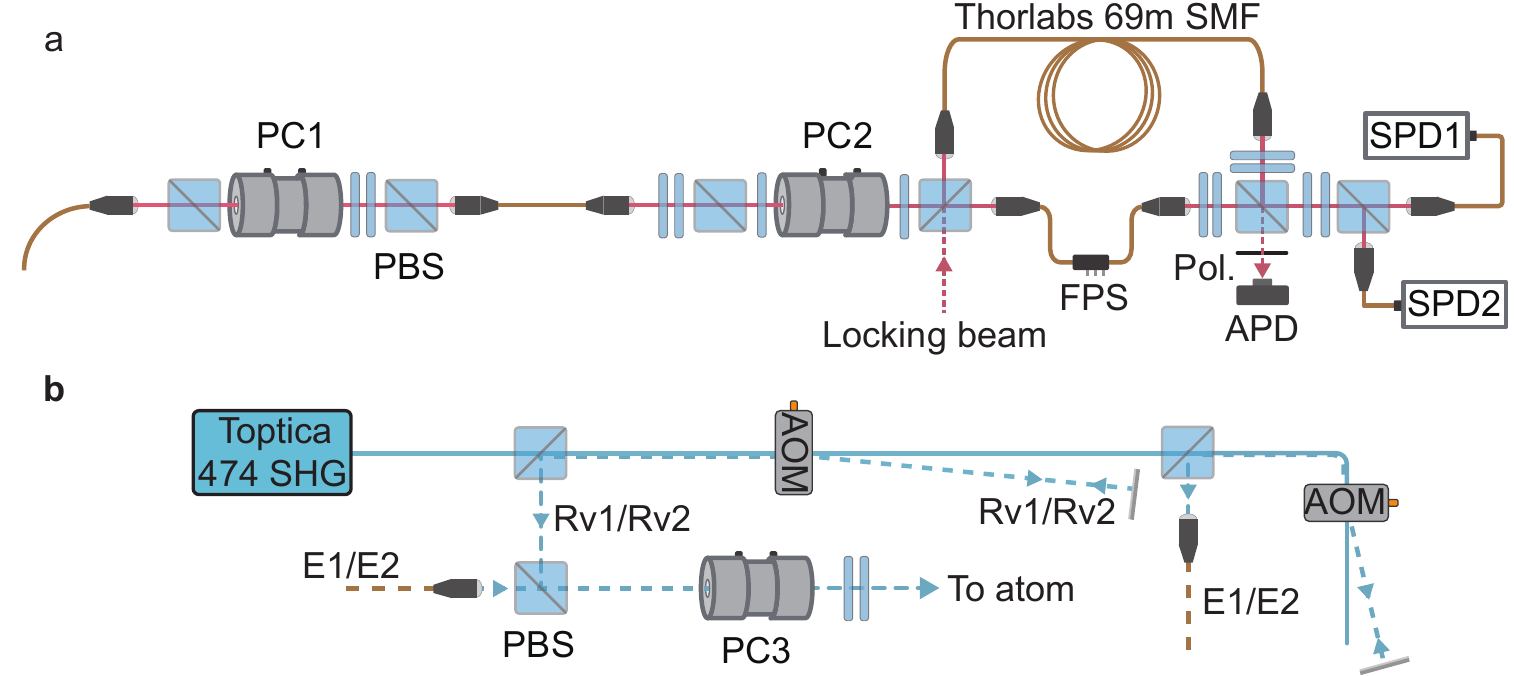}
	\caption{\textbf{More details on experimental setup.} \textbf{a}, Measurement setup. PC: Pockels cell, SMF: Single mode fiber, APD: Avalanche photodiode, PBS: Polarizing beam splitter, SPD: Single photon detector. \textbf{b}, 474 nm laser system (some waveplates not shown).}
	\label{fig:setup}
\end{figure}

The measurement apparatus is shown in Fig.~\ref{fig:setup}a. The Pockels cell of PC1 and the adjoining PBSs are used to filter the noise of excitation pulses in the temporal domain by rapidly switching the noise pulses to the orthogonal polarization and filtering with a PBS. The Pockels cells are driven by high-voltage pulsed drivers. Then PC2 is used to switch the early mode to the long fiber of 69 m and the late mode to the short fiber of 5 m. We use a weak locking laser beam coupled through another port of the PBS to lock the interferometer. The locking beam leaves the interferometer from another port to the APD. Then the interferometer is locked on the error signal generated from the modulated locking beam. We dynamically turn off the locking beam in the experimental period to prevent residual noise from leaking to the SPDs in the detection windows.

We use the non-polarization maintaining fibers. The polarizations of the single photon and the locking beam are orthogonal in the fiber. We consider the drift of the birefringence effect of the fiber is possible to change the relation between two polarizations in long period and thus
the locking point, so we cover the long fiber with thermal insulation material. The fiber phase shifter (FPS) with a fiber length of 5 m works as the short arm of the interferometer and also for phase feedback. The FPS works by stretching the fiber convolved on a small PZT, which may also induce a birefringence phase in the fiber. To avoid the possible change of the birefringence phase in the long time, we stabilized the temperature of the FPS.

The superatom is manipulated by a SHG (Second-Harmonic Generation) laser of 474 nm from Toptica with a power of $\sim$1.1W. As shown in Fig.~\ref{fig:setup}b, the laser is frequency shifted with an AOM, and the zero-order beam is reused to generate the excitation pulses (E1/E2). The AOMs are drived by a dual-channel arbitrary function generator (AFG). The excitation and retrieval pulses are driven by two synchronized channels of the AFG and combined with a PBS, and PC3 is used to switch the corresponding polarizations. The maximum power of $Rv_2$ on the superatom is about 500 mW with
a waist radius of 19 $\upmu$m. In addition, the seed laser of 948 nm and another laser of 795 nm are both locked on an ULE (Ultra-Low Expansion) cavity.

\section{Experimental time sequence}

\begin{figure}[hbtp]	
	\centering
	\includegraphics[width=0.6\textwidth]{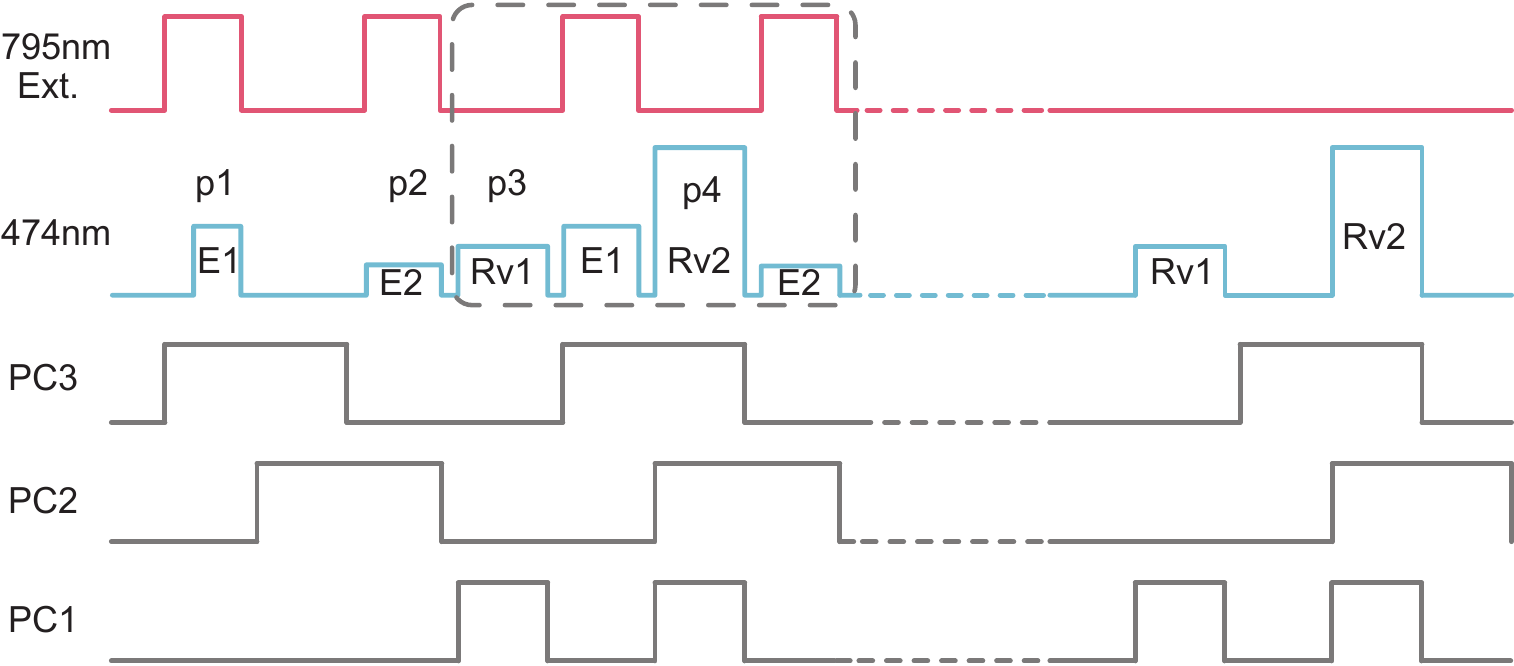}
	\caption{\textbf{Time sequence.} The 474 nm waveforms are modified with the AFG. Other time sequences are generated with the FPGA. The pulses marked as p1 to p4  correspond to the preparation and retrieval of the first photon. The pules in the dashed box are iterated to generate more photons.}
	\label{fig:timeseq}
\end{figure}

The experimental time sequences are shown in Fig.~\ref{fig:timeseq} and generated with an FPGA system. We generate the excitation pulses of 795 nm (Ext.) and 474 nm (E1/E2) with a width of 100 ns, together with PC3 controlling the polarizations of the 474 nm pulses. The width of the first excitation pulse is narrowed to realize a $\pi/2$ excitation. Then together with another patching pulse of $\ket{R_2}$, the initial preparation is finished. The $Rv_1/Rv_2$ pulses with the width of 170 ns then retrieve the entangled single photons, and the patching pulses (E1, E2 and 795nm Ext.) are used to recreate the qubits. The power, width and the frequency of the 474 nm pulses are all modified with the AFG, and we also sweep the relative phase of the GHZ states by changing the waveforms. Another Pockels cell of PC2 dynamically bring backwards the late photons retrieved with $Rv_2$ through the long fiber to overlap with the early photons. The noises from the $\ket{R_2}$ excitation are also brought backwards by PC2. Together with the filtering of PC1, the detection probability of noise in each SPD and each window is reduced to about 0.1\%. Besides, the 795 nm excitation beam is resonant with the ring cavity of the superatom, so the power requirement is reduced a lot, which can also help reduce the noise going into the SPDs.

\section{Fixing the internal phase }

To prepare the GHZ states, we first measure the oscillations of the coincidences in the $(\ket{E}\pm\ket{L})/\sqrt{2}$ basis by varying the relative phases with AFG for each $\ket{\Phi}_m$ state. We decompose $\ket{\Phi}_m$ with $\ket{D}=(\ket{E}+\ket{L})/\sqrt{2}$ and $\ket{A}=(\ket{E}-\ket{L})/\sqrt{2}$:

\begin{equation*}
	\begin{split}
	\ket{\Phi}_m^\phi=&\frac{1}{\sqrt{2}} (\ket{E}^{\otimes m}+e^{i\phi}\ket{L}^{\otimes m})\\
	=&\sum_{s=0,2,4...}\sum_{j_1,...,k_1,...} \ket{D}^{\otimes (m-s)}_{j_1j_2...}\ket{A}^{\otimes s}_{k_1k_2...} (1+e^{i\phi})/2^{(m+1)/2}\\
	+&\sum_{s=1,3,5...}\sum_{j_1,...,k_1,...} \ket{D}^{\otimes (m-s)}_{j_1j_2...}\ket{A}^{\otimes s}_{k_1k_2...} (1-e^{i\phi})/2^{(m+1)/2}.
	\end{split}
\end{equation*}
The symbols of $j_1,j_2,...$  and $k_1,k_2,...$ are the photon indexes with the conditions of $j_1<j_2<j_3...$ and $k_1<k_2<k_3...$ . So the coincidences of $2^m$ terms are separated to two groups both of which are summed up. So we can determine the relative phase of zero from the normalized oscillations with $\phi$ (not shown), and $\phi$ is swept by changing the initial phase of $Rv_2$.

\section{Original data for $\ket{\Phi}_1$ and $\ket{\Phi}_2$}

After setting the relative phase, we measure the coincidence in different bases. Each photon detected can provide one count of SPD$_1$ or SPD$_2$, corresponding to the orthogonal basis. Then we can get all coincidences of $2^m$ terms shown in the histograms of Fig.~2a. For $\ket{\Phi}_1$ the counts of $\ket{E}\bra{E}$ and $\ket{L}\bra{L}$ are 13724 and 15240, and the counts of $\ket{D}\bra{D}$ and $\ket{A}\bra{A}$ are 28097 and 642. The experimental cycle numbers of two settings are both about $2.6\times10^5$ measured in 40 s.

\begin{table*}[hbtp]	
	\centering
	\caption{Coincidence of $\ket{\Phi}_2$ on eigenstate and superposition state basis}
	\begin{threeparttable}
		\begin{tabular}{cccccc}
			\toprule
			\hline
			Basis& \quad\quad & \quad\quad$EE$ \quad\quad& \quad\quad$EL$\quad\quad & \quad\quad$LE$\quad\quad & \quad\quad$LL$\quad\quad \\
			Coincidences& \quad\quad & \quad\quad2407 \quad\quad& \quad\quad154\quad\quad & \quad\quad71\quad\quad & \quad\quad2301\quad\quad \\
			\midrule
			Basis& \quad\quad & \quad\quad$DD$ \quad\quad& \quad\quad$DA$\quad\quad & \quad\quad$AD$\quad\quad & \quad\quad$AA$\quad\quad \\
			Coincidences& \quad\quad & \quad\quad2234 \quad\quad& \quad\quad179\quad\quad & \quad\quad204\quad\quad & \quad\quad2324\quad\quad \\
			\midrule
			Basis& \quad\quad & \quad\quad$CC$ \quad\quad& \quad\quad$CP$\quad\quad & \quad\quad$PC$\quad\quad & \quad\quad$PP$\quad\quad \\
			Coincidences& \quad\quad & \quad\quad225 \quad\quad& \quad\quad2237\quad\quad & \quad\quad2286\quad\quad & \quad\quad192\quad\quad \\
			\hline
			\bottomrule
		\end{tabular}
	\end{threeparttable}
	\label{fig:phi2}
\end{table*}

For $\ket{\Phi}_2$ state, the detailed results are shown in Table.~\ref{fig:phi2} with $\ket{C}=(\ket{E}+i\ket{L})/\sqrt{2}$ and $\ket{P}=(\ket{E}-i\ket{L})/\sqrt{2}$. The bases in the table are written in a simplified form. The experimental cycle number of each row is about $4\times10^5$ measured in 60 s. According to the definition in the main text, we can calculate the fidelity with the coincidences:
\begin{gather*}
F_e= (c_{EE}+c_{LL})/(c_{EE}+c_{EL}+c_{LE}+c_{LL}),\\
M_0^{\otimes 2} = (c_{DD}-c_{DA}-c_{AD}+c_{AA})/(c_{DD}+c_{DA}+c_{AD}+c_{AA}),\\
M_1^{\otimes 2} = (c_{CC}-c_{CP}-c_{PC}+c_{PP})/(c_{CC}+c_{CP}+c_{PC}+c_{PP}),\\
F_s =  (M_0^{\otimes 2}-M_1^{\otimes 2})/2.
\end{gather*}
Besides, the probability ratio of single photon to noise is about 100 : 1. So the noise can contribute to the coincidences a little. The noise is not corrected in the main text. We estimate $c_{EL}\approx 107$ and $c_{EL}\approx 24$ with the noise corrected.

\section{Temporal profile analysis}

\begin{figure}[hbtp]	
	\centering
	\includegraphics[width=0.6\textwidth]{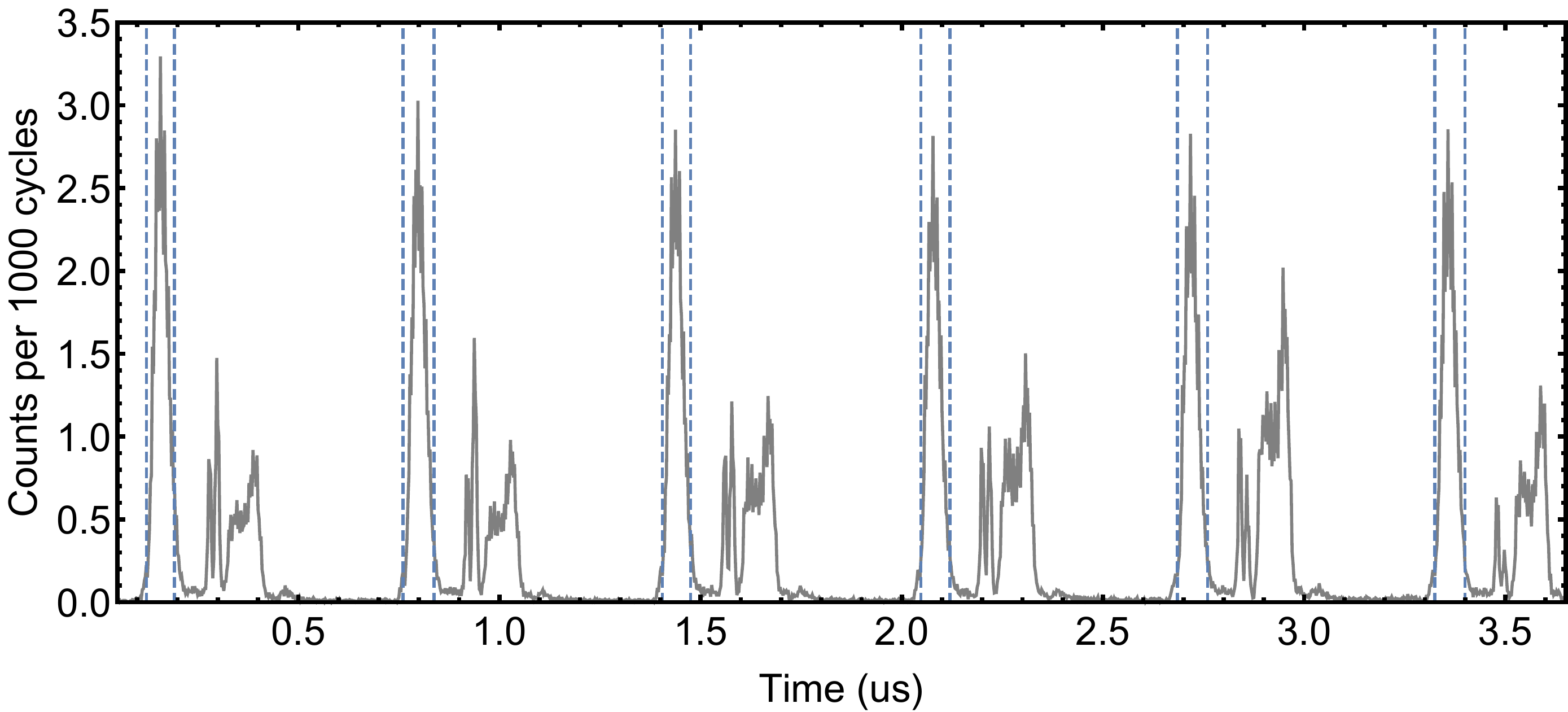}
	\caption{\textbf{Profile of the $\ket{\Phi}_6$ photons.} The measurement resolution of each data point is 2.5 ns. The detection windows are defined by the dashed lines. The residual noise is mainly after each detection window. The profile detected with SPD$_2$ is similar and not shown. We sweep the phase of the interferometer, so the results of two SPDs can be averaged.}
	\label{fig:phwave}
\end{figure}

The profiles of the single photon pulses are shown in Fig.~\ref{fig:phwave}, with the dashed lines defining the detection windows. The time gap between adjacent windows is 640 ns. Residual excitation noises occur after each detection window. The position of the noise is manipulated by PC2. Because the switching time of the Pockels cell is within tens of nanoseconds, some signals and noises at the edge are not fully switched. We set the window narrower than the signal which seems to have slight improvement to the fidelity, and this also contributes to the efficiency scaling factor. In each window, the second-order auto-correlation function $g^{(2)}$ is measured to be $\sim 0.05$, mainly limited by the dark counts of 0.1\% ($g^{(2)}<$0.01 after correcting the dark counts). The value of $g^{(2)}=p_{s1s2}/(p_{s1} p_{s2})$ is measured by the Hanbury Brown-Twiss experiment, with $p_{s1s2}$ referring to the probability of coincidence and $p_{s1/s2}$ referring to the individual probability of counts on SPD$_{1/2}$.

From the profiles, we find the detected efficiency in the first window corresponds to the overall single photon efficiency of 9.4\%. But the heights of the signals in the following windows gradually decrease. To find out the problem, we analyse the original data of Fig.~2a. We calculate the conditional probability of $P_{(i+1)|i}$, which refers to the probability of detecting the photon in the window $i+1$ conditioned on the detection in the window $i$, and find that the conditional probabilities are not decreasing with the increase of $i$.  So the in-fiber efficiency does not show obvious dropping, and can not be directly
associated with the phenomenon in the time domain. Thus we infer each patching operation has a chance to generate some accumulated components in the Rydberg state, which will suppress subsequent excitations and can not be efficiently retrieved during the successive photon generation. As a result, the integral probability of the accumulated components will increase and thus the photon counts will decrease corresponding to the time domain. We think the accumulated components possibly come from the residual double Rydberg excitation due to our fast excitation of 100 ns. In the end of each cycle, we have already applied additional $Rv_1$ and $Rv_2$ pulses of hundreds of nanoseconds, which may help clean the accumulated components. From the dropping speed in Fig.~\ref{fig:phwave}, the probability of creating the accumulated components for each patching operation is estimated to be about 2\%. Once the accumulated components are created, the detection of successive photons will fail. Therefore the fidelity will not be influenced.

\section{Bit-flip error}

\subsection{Patching pulse infidelity}

Some issues will cause the imperfection of the patching pulse, such as the fluctuation of the collective Rabi frequency coming from the passive stability of atom numbers or laser power. Two retrieving and two patching pulses are used to expand one photon and recreate the superatom qubits. Initialized on $\ket{1}_1\ket{0}_2$, the failure of the first patching pulse will cause the flip from $\ket{1}_1$ to $\ket{0}_1$, and thus the second qubit will flip from $\ket{0}_2$ to $\ket{1}_2$. Also, for the $\ket{0}_1\ket{1}_2$ state, the failure of the second patching pulse will result in the $\ket{0}_1\ket{0}_2$ state and eliminate the successive generation.

\begin{figure}[hbtp]	
	\centering
	\includegraphics[width=0.45\textwidth]{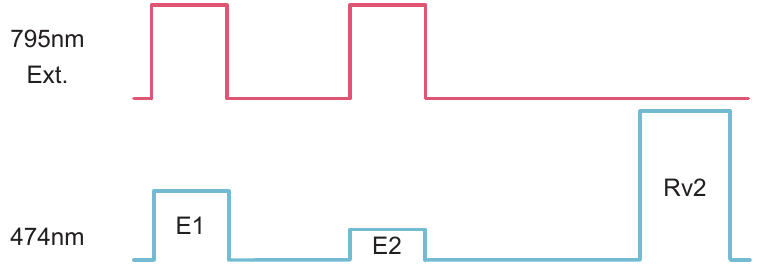}
	\caption{\textbf{Independent measurements of $\ket{R_1}$ patching infidelity.} }
	\label{fig:indepmeas}
\end{figure}

In order to observe the failure of the patching pulse, we perform an independent measurement as shown in Fig.~\ref{fig:indepmeas}. We remove other operations and simply use $\ket{R_{1}}$ created with the patching pulse to suppress the subsequent $\ket{R_{2}}$. We find that compared with the situation with $\ket{R_{1}}$ not created, the photon retrieved from $\ket{R_{2}}$ is suppressed to about 2\%. We think the residual signal can be mainly explained as the failure of the $\ket{R_{1}}$ patching pulse. With a similar experiment, the failure of the $\ket{R_{2}}$ patching pulse is also measured to be about 2\%.

\subsection{Unexpected retrieval}

\begin{figure}[hbtp]	
	\centering
	\includegraphics[width=0.45\textwidth]{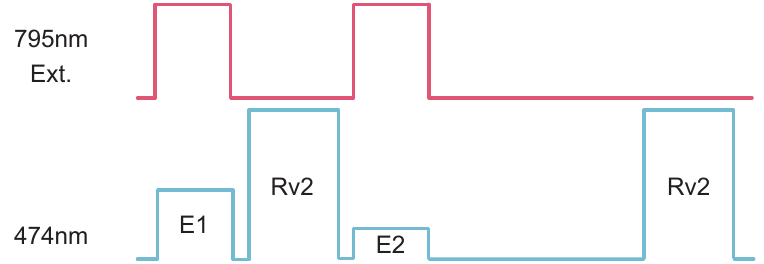}
	\caption{\textbf{Independent measurements of the unexpected retrieval of $Rv_2$.} }
	\label{fig:indepmeas1}
\end{figure}

Further, we insert $Rv_{2}$ pulse between the excitations of $\ket{R_{1}}$ and $\ket{R_{2}}$ as shown in Fig.~\ref{fig:indepmeas1}, and find that the retrieval pulse will cause the residual signal from $\ket{R_{2}}$ to increase. This can be explained as the unexpected retrieval of $\ket{R_{1}}$. The polarization impurity of the retrieving pulses and finite splitting of the Rydberg energy levels can lead to the loss of $\ket{R_{1}}$. From the residual signal, we estimate that
$\ket{R_{1}}$ will experience unexpected retrieval by $Rv_{2}$ with a probability of about 3\%. With a similar experiment, we estimate $\ket{R_{2}}$ will experience unexpected retrieval by $Rv_{1}$ with a probability of about 1\%. We think the distinction comes from the the imbalance of the dipole matrix elements ($\bra{e_1}er\ket{r_1}/\bra{e_1}er\ket{r_2}=\sqrt{3}$)  and the intensity of the retrieving pulses ($I_{R_1}/I_{R_2}\sim1/3$). Thus the $\ket{R_1}$ state is more sensitive to the polarization impurity.

\subsection{Eigenstate fidelity}
\label{bitflip}
In each step, we cycle retrieval-patching-retrieval-patching operations to expand one photon and the superatom qubit. But as mentioned above,  each operation may fail with small probabilities. In Table.~\ref{fig:bitflip}, we list the complete evolution when expanding one photon, with a failure of one of the operations. There we do not consider the failure of more than one operation because the high-order events have much lower probability.

\begin{table*}[hbtp]	
\centering
\caption{Evolution of the atomic state during each iteration}
\begin{threeparttable}
	\begin{tabular}{cccccccccccccc}
		\toprule
		atomic state  & \multicolumn{4}{c}{$\ket{0}_1\ket{1}_2$}  & \quad & \quad & \quad & \multicolumn{4}{c}{$\ket{1}_1\ket{0}_2$} & \quad & \quad \\
		steps\quad\quad & $Rv_1$ & $P_1$ & $Rv_2$ & $P_2$ &\quad error\quad&  prob. \quad & \quad & $Rv_1$ & $P_1$  & $Rv_2$ & $P_2$ &  \quad error \quad& prob.\quad \\
		\midrule
		$Rv_1$ fails  & $\widetilde{00}$ & 10 & 10 & 10 &\quad flip & 1\% & \quad & $\widetilde{00}$ & 10 & 10 & 10 & \quad  & \quad \\
		$P_1$ fails  & 01 & 01 & $\widetilde{00}$ & 01 & \quad  & \quad & \quad & $\widetilde{00}$ & 00 & 00 & 01 & \quad flip  & 2\%   \\
		$Rv_2$ fails & 01 & 01 & $\widetilde{00}$ & 01 & \quad  & \quad & \quad & $\widetilde{00}$ & 10 & $\widetilde{00}$ & 01 & \quad flip & 3\%  \\
		$P_2$ fails & 01 & 01 & $\widetilde{00}$ & 00 & \quad loss  & 2\% & \quad & $\widetilde{00}$ & 10 & 10 & 10 & \quad & \quad  \\
		\bottomrule
	\end{tabular}
	\begin{tablenotes}
		\footnotesize
		\item[1]  The failure of $Rv_{1/2}$ is defined as that the operation succeeds to retrieve $R_{1/2}$ but meanwhile also retrieves $R_{2/1}$. The failure of $P_{1/2}$ means the $\pi$ pulse on $R_{1/2}$ fails to create an excitation.
		\item[2]  The tilde overline means the operation can lead to the single photon releasing.
		\item[3] Prob. refer to the probability for the single failure to happen.
	\end{tablenotes}
\end{threeparttable}
\label{fig:bitflip}
\end{table*}

Beginning with the superatom state of $\ket{0}_1\ket{1}_2$ or $\ket{1}_1\ket{0}_2$, when we implement $Rv_1\rightarrow P_1\rightarrow Rv_2\rightarrow P_2$ operations, errors may happen with either one of them failing. The $\ket{0}_1\ket{1}_2$ state will flip with the unexpected retrieval happening. Nevertheless, the loss event will not contribute to the m-fold coincidence.  The $\ket{1}_1\ket{0}_2$ state can flip in two situations, and result in much more contributions to the infidelity. The difference
between $\ket{0}_1\ket{1}_2$ and $\ket{1}_1\ket{0}_2$ comes from the sequential order of our time sequence.

 We define $p_1$ as the infidelity on $\ket{0}_1\ket{1}_2$ and $p_2$ as the the infidelity on $\ket{1}_1\ket{0}_2$. From Table.~\ref{fig:bitflip} we can calculate $p_1\approx 1\%$ and $p_2\approx 8\%$. There the $Rv_2$ failure on $\ket{1}_1\ket{0}_2$ can lead to the release of another photon, and thus we simply consider its contribution to infidelity to be doubled. The averaged probability for the errors happening on $\ket{0}_1\ket{1}_2$ and $\ket{1}_1\ket{0}_2$  is $(p_1+p_2)/2\approx 4.5\%$. We can simply estimate the scaling of $F_e$ being $0.955^{m-1}$. So the independent measurements give a close result compared with the fitting curve of $\alpha^{m-1}=0.96^{m-1}$ in the main text.

\subsection{Relation between $F_e$ and $F_s$}\label{rela}

As discussed above, the errors in each step can cause the generation of some unexpected quantum states. We simply consider the patching failure in step s and lead to the flipped state expressed as $\ket{\varepsilon_s}$. We assume there are no other additional dephasing terms. This process can be expressed as
$$\ket{\Phi}_m \Rightarrow \ket{\Phi'}_m = q_1 \ket{E}^{\otimes m}+q_2\ket{L}^{\otimes m} +\sum k_s e^{i \phi_s}\ket{\varepsilon_s}$$
with $q_1, q_2>0$. It is easy to derive the fidelity of $F=|\langle \Phi'_m\ket{\Phi_m}|^2=(q_1+q_2)^2/2$.  Since $\langle \varepsilon_s\ket{\Phi_m}$=0, $\ket{\varepsilon_s}$ and the phase $\phi_s$ will not contribute to the fidelity. Then $F_e$ can also be easily calculated with $q_1^2+q_2^2$ from the definition. Then we use an easy way to derive $F_s=2 F-F_e=2 q_1 q_2$, and thus $F_s\leqslant F_e$, which shows $F_e$ is the upper limit of $F_s$. In most situations, the quantum states are balanced, so we can consider $q_1^2\approx q_2^2$ and $F_e\approx F_s$. If there are phase fluctuations between $\ket{E}^{\otimes m}$ and $\ket{L}^{\otimes m}$, $F_s$ will be further reduced.

\section{Other issues on the eigenbasis result}

\subsection{Imbalance of the eigenstate}

We find that the 3D histogram in the main text gradually shows imbalance with the increasing of photon number. And the imbalance of the $\ket{\Phi}_6$ state is pretty obvious. The imbalance of the bit-flip error can partly explain this with the flip of $\ket{E}$ being relatively larger and accumulated over m photons. Also, we find the first photon shows some imbalance which maybe comes from the drift of the first $\pi$/2 pulse after long time experiment. Considering these two reasons, The count of of $\ket{E}\bra{E}^{\otimes m}$ is estimated to be about 0.7 times of $\ket{L}\bra{L}^{\otimes m}$. Further, we think the statistical fluctuation of Poisson distribution also contributes a lot to the imbalance considering the error bar is large.

\subsection{Influence of the detector afterpulse}

The SPDs have a chance to give the afterpulse which means it can give another fake count in the next window after detecting the real photon. And it will cause the fake coincidences. Between the adjacent windows of 640 ns, the afterpulse of the SPDs in this work is measured to be about 0.1\%. In a simple situation, if there are some 3-fold event of $E0LL$ with the second photon not detected, then the afterpulse will cause additional fake $EELL$ coincidence with the probability of 0.1\%. We therefore consider all the possible events and correct the experimental result, then the fidelity of $\ket{\Phi}_6$ is calculated to be 62.8\%. Considering the influence of the afterpulse is not too much, we do not correct it in the main text.

\section{Phase noise measurement}

Firstly, the fidelities are influenced by the performance of the measurement devices, such as the precision of the phase locking. We use the weak probe laser of the 795 nm and the oscilloscope to test the locked measurement devices. The standard deviation $\sigma_{inter}$ of phase fluctuation is measured to be about 7.4$^\circ$ calculated from the intensity signal of interference. Further, although the laser is locked on the ULE cavity, there is residual phase noise of around megahertz which will also influence the fidelity~\cite{de2018analysis}.

\begin{figure}[hbtp]	
	\centering
	\includegraphics[width=0.4\textwidth]{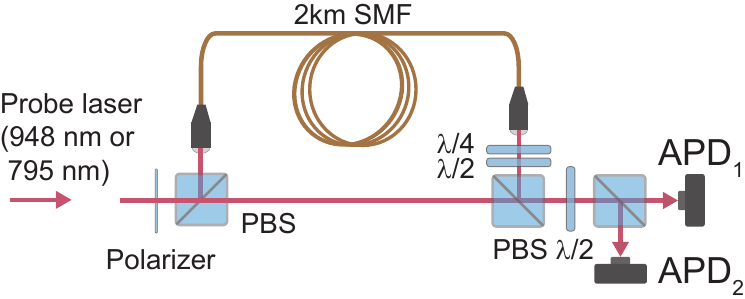}
	\caption{\textbf{Measurement of the laser phase noise.} The relative phase of two arms is not locked, as the drift of the interferometer is slow. The phase fluctuation of the laser can lead to fast variation of the intensity in several microseconds, measured by two APDs.  }
	\label{fig:phasmea}
\end{figure}

We analysed the laser noise with a 10 $\upmu$s-delayed  Mach-Zehnder interferometer as shown in Fig.~\ref{fig:phasmea}. We split the laser power to two parts with a PBS and let them interfere with each other but with one arm delayed 10 $\upmu$s. With infinite length of delaying fiber, the process is equivalent to the interference of two independent laser locked on the same frequency, so the phase correlation between them will be eliminated and the intensity fluctuation in the temporal domain can indicate their phase noise of $\sqrt{2}\sigma_{laser}$, and $\sigma_{laser}$ is the standard deviation of one laser pulse. In this work, the time interval between adjacent operations is hundreds of nanoseconds, so we think the delay of 10 $\upmu$s is sufficient. We measure the intensity fluctuation with two APDs and subsequently calculate the phase evolution $\varphi(\tau)$. By directly calculating $\varphi(\tau_0)-\varphi(\tau_0+\Delta \tau)$, we can analyse its influence on the superatom at different time $t_0$. Then we calculate the standard deviation of $\sigma_{laser_1}$ being $ 4.8^\circ$ with $\Delta \tau$= 480 ns for the 795 nm laser.

 The phase fluctuation of the 474 nm laser comes from two step joint operations of creating one photon as shown in Fig.~\ref{fig:timeseq}. The phases of the p1 to p4 pulses can be overlapped with the laser phase noise of $\varphi_{p1}$ to $\varphi_{p4}$. We thus analyse its influence of $\Delta\varphi(\tau_0)=(\varphi_{p1}-\varphi_{p3})-(\varphi_{p2}-\varphi_{p4})$ when creating a single photon. The result gives a value of $\sigma_{laser_2}$ being about $ 12^\circ$. So the standard deviation of the overall laser noise is therefore $\sqrt{\sigma_{laser_1}^2+\sigma_{laser_2}^2}\approx 13^\circ$  Also, we extend the analysis of data to more photons, and find that the noise increased to $\sqrt{m}$ times which indicates that the laser noises on different photons have no obvious correlation.

\section{Phase-fidelity relation}

In Sec.~\ref{rela}, we analyse the simple situation without additional dephasing. Further, we assume there is a relative phase between $\ket{E}^{\otimes m}$ and $\ket{L}^{\otimes m}$, and the phase noise is described by the normal distribution of $p(\phi)=e^{-\phi^2/(2 \sigma_r^2)}/\sqrt{2\pi \sigma_r^2}$. Here $\sigma_r$ is the standard deviation of the phase noise in radian. It is easy to derive the fidelity $F=\int_\phi(q_1^2+q_2^2+2q_1q_2 {\rm cos}\phi)d\phi$ and $F_s=\int_\phi 2 q_1q_2 {\rm cos}(\phi)d\phi$. We find $F_s=2q_1q_2 e^{-\sigma_r^2/2}\approx F_e e^{-\sigma_r^2/2}$ with the dephasing term in the form of $e^{-\sigma_r^2/2}$. The $m$-photon infidelity
results from all above noise of $\sqrt{m^2\sigma^2_{inter}+m\sigma^2_{laser}}$. There, the fluctuation of the time-bin locking point is slow and the drift is the same for all $m$ photons, so the noise of $m$-fold rapidly increases to $m\sigma_{Lock}$. Then we calculate $F_s$ with the relation of multiplication of different contributions expressed as $F_s=F_e \beta_\phi \beta_{res}$ with $\beta_\phi=\beta_{inter}^{m^2}\beta_{laser}^m$. $\beta_{laser}$ and $\beta_{inter}$ are the dephasing terms of one photon.

\end{document}